\documentclass[11pt]{llncs}
\usepackage{multirow}
\usepackage{enumitem}
\usepackage{amssymb}
\usepackage{amsmath}
\usepackage{mathtools}

\begin{document}

\title{On the relationship between (secure) multi-party computation and (secure) federated learning}
\author{Huafei Zhu}
\institute{IHPC, A*STAR, Singapore \\ zhu\_huafei@ihpc.a-star.edu.sg}

\maketitle
\begin{abstract}
The contribution of this short note, contains the following two parts:
\begin{itemize}

\item in the first part, we are able to show that the federate learning (FL) procedure presented by Kairouz et al.~\cite{Kairouz1901}, is a random processing. Namely, an $m$-ary functionality for the FL procedure can be defined in the context of multi-party computation (MPC); Furthermore, an instance of FL protocol along Kairouz et al.'s definition can be viewed as an implementation of the defined $m$-ary functionality. As such, an instance of FL procedure is also an instance of MPC protocol. In short, FL is a subset of MPC.

\item To privately computing the defined FL (m-ary) functionality, various techniques such as homomorphic encryption (HE), secure multi-party computation (SMPC) and differential privacy (DP) have been deployed. In the second part, we are able to show that if the underlying FL instance privately computes the defined $m$-ary functionality in the simulation-based framework, then the simulation-based FL solution is also an instance of SMPC. Consequently, SFL within the simulation framework is a subset of SMPC.
\end{itemize}
\end{abstract}

\section{Introduction}
Since the concept of federated learning (FL) first introduced by McMahan et al.~\cite{Mahan1701}, numerous user data protection solutions leveraging various security mechanisms such as secure multi-party computation (say, \cite{Dan2012,Mahan1702}), homomorphic encryption (say, \cite{Hardy1701,CL2001}) and differential privacy (say, \cite{Kangwei2001}) have been investigated and published. We refer to the reader \cite{He1901,YQ1901,Kairouz1901} and references therein for further reading. To the best of our knowledge, the relationship between multi-party computation (MPC) and federated learning (FL) as well as that of secure multi-party computation (SMPC) and secure federated learning (SFL) have not been well addressed. This note aims to provide an initial insight to the relationship on (S)MPC vs.(S)FL and we are able to show that 
\begin{itemize}
\item FL is a subset of MPC;

\item If FL attains the security in the simulation-based framework, then SFL is a subset of SMPC.
\end{itemize}

\section{Preliminaries}
 
\subsection{$m$-ary functionality, MPC and SMPC}
We briefly describe the notations and notions of $m$-ary functionality, MPC and SMPC below and refer to the reader~\cite{Lindell201601,Goldreichbook1,Goldreichbook2} for more details.

\subsubsection{$m$-ary functionality} 

An $m$-ary functionality, denoted by $f$: $(\{0, 1\}^*)^m$ $\rightarrow$ $(\{0, 1\}^*)^m$, is a random process mapping string sequences of the form $\overline{x}$ = $(x_1$, $\cdots$, $x_m)$ into sequences of random variables, $f_1(\overline{x})$, $\cdots$, $f_m(\overline{x})$ such that, for every $i$, the $i$th party $P_i$ who initially holds an input $x_i$, wishes to obtain the $i$th element in $f(x_1, \cdots, x_m)$ which is denoted by $f_i(x_1,\cdots, x_m)$.

\subsubsection{Multi-party computation} 
A multi-party computation (MPC) problem is casted by specifying an implementation of the defined $m$-ary functionality. Namely, an MPC protocol is a procedure computing the defined $m$-ary functionality. We emphasize that the notion of MPC does NOT guarantee the proposed MPC protocol securely computing the defined m-ary functionality. It is possible where no security is introduced in the MPC protocol at all. 

\subsubsection{Secure multi-party computation} A multi-party computation securely computes an $m$-ary functionality (i.e., secure multi-party computation, SMPC) if the following security definition is satisfied.

Let $[m]$ = $\{1$, $\cdots$, $m\}$. For $I \in \{i_1, \cdots, i_t\}$ $\subseteq$ $[m]$, we let $f_I(x_1, \cdots, x_m)$ denote the subsequence $f_{i_1}(x_1, \cdots, x_m)$, $\cdots$, $f_{i_t}(x_1, \cdots, x_m)$. Let $\mathrm{\Pi}$ be an $m$-party protocol for computing $f$. The view of the $i$-th party during an execution of $\mathrm{\Pi}$ on $\overline{x}$:= $(x_1, \cdots, x_m)$ is denoted by $\mathrm{View_i ^{\Pi}} (\overline{x})$. For $I$ = $\{i_1, \cdots, i_t\}$, we let $\mathrm{View_I ^{\Pi}} (\overline{x})$:= ($I$, $\mathrm{View_{i_1} ^{\Pi}} (\overline{x})$, $\cdots$, $\mathrm{View_{i_t} ^{\Pi}} (\overline{x})$). 

\begin{itemize}
\item In case $f$ is a deterministic $m$-ary functionality, we say $\mathrm{\Pi}$ privately computes $f$ if there exists a probabilistic polynomial-time algorithm denoted $S$, such that for every $I \subseteq [m]$, it holds that $S(I$, $(x_{i_1}, \cdots, x_{i_t})$, $f_I(\overline{x}))$ is computationally indistinguishable with $\mathrm{View_I ^{\Pi}} (\overline{x})$. 

\item In general case, $S(I$, $(x_{i_1}, \cdots, x_{i_t})$, $f_I()$, $f(\overline{x}))$ is computationally indistinguishable with $\mathrm{View_I ^{\Pi}}$ $((\overline{x})$, $f(\overline{x}))$.
\end{itemize}

An oracle-aided protocol is a protocol augmented by a pair of oracle types, per each party. An oracle-call step is defined as follows: a party writes an oracle request on its own oracle tape and then sends it to the other parties; in response, each of the other parties writes its query on its own oracle tape and responds to the first party with an oracle call message; at this point the oracle is invoked and the oracle answer is written by the oracle on the ready-only oracle tape of each party. An oracle-aided protocol is said to privately reduce $g$ to $f$ if it securely computes $g$ when using the oracle-functionality $f$. In such a case, we say that $g$ is securely reducible to $f$.

\subsection{Federated Learning Process}
A federated learning server (FLS) orchestrates the training process, by repeating the following steps until training is stopped. \\

\noindent \framebox{\parbox[c]{12.5cm}{\center{\underline{Federated learning procedure~\cite{Kairouz1901} } }
\begin{enumerate}
\item Client selection: The server specifies $(m-1)$-client meeting eligibility requirements;
\item Broadcast: The selected clients download the current model weights and a training program from the server. The model weights and training program are part of system parameters;
\item Client computation: Each selected client locally computes an update to the model by executing the training program;
\item Aggregation: The server collects an aggregate of the clients' updates. If no privacy requirement is introduced, FL will be called plain-FL (or FL for short, in this short note).

\item Model update: The server locally updates the shared model based on the aggregated update computed from the clients that participated in the current round.
\end{enumerate}
}}

\section{The relationship between MPC and FL and SMPC and SFL}
For each iteration $j$ defined in the FL procedure, we are able to define an $m$-ary functionality $f_j$ for the current round. With the help of the oracle-aided reduction, we know that FL functionality $f_{\rm {FL}}$ can be defined as a composition $f_n \circ f_{n-1} \circ \cdots \circ f_1$ of the round functionalities, where $f_j$ is the $m$-ary functionality for iteration $j$. As such, for each iteration $j$, if there exists an $m$-party protocol privately computing $f_j$, then by applying the SMPC composition theorem\cite{Lindell201601,Goldreichbook1,Goldreichbook2}, we are able to show there is an $m$-party protocol privately computing $f_{\rm {FL}}$. To complete this work, we need to define an $m$-ary functionality for each iteration of the FL procedure. The details of the $m$-ary functionality are depicted below.  \\

\noindent \framebox{\parbox[c]{12.5cm}{\center{\underline{For each iteration in FL, we define an m-ary functionality: } }
\begin{enumerate}
\item Define $m$-1 parties selected by the FL-server (FLS) as MPC participants. Including FLS itself, there are $m$-party;
\item Define the model weights and training program as system parameter ($sysparam$) to MPC;
\item Define initial $View_i$ as input $D_i$, randomness $r_i$ and $sysparam$ for $P_i$. $View_i$ is append-only data type.  

\item Define $f_m$: $View_{\rm{FLS}}$ $\rightarrow$ ($sout_1$, $\cdots$, $sout_m$), where $sout_i$ denotes server's output such that $P_i$ gets its output $sout_i$ ($i$ = 1, $\cdots$, $m-1$). The output of FLS is $sout_m$.

\item Define $m$-ary $f$ = ($f_1$, $\cdots$, $f_m$), where $f_i$: $view_i$ $\rightarrow$ $sout_i$.
\end{enumerate}
}}

\section{Summary}

Putting the above discussions together, we get two claims explicitly stated in the abstract.

\end{document}